\documentclass[letterpaper, 10 pt, conference]{ieeeconf}  

\IEEEoverridecommandlockouts                              

\usepackage{graphics} 
\usepackage{graphicx}
\usepackage{subfigure}
\usepackage{epsfig} 
\usepackage{mathptmx} 
\usepackage{times} 
\usepackage{amsmath} 
\usepackage[T1]{fontenc}
\usepackage{amsfonts,amssymb}
\usepackage{bm}

\usepackage{CJK}
\usepackage{algorithm}
\usepackage{algorithmicx}
\usepackage{algpseudocode}
\newcommand{\RNum}[1]{\uppercase\expandafter{\romannumeral #1\relax}}

\usepackage{soul} 
\usepackage{color, xcolor} 
\usepackage{url} 
\usepackage{soul}
\soulregister\cite7 
\soulregister\citep7 
\soulregister\citet7 
\soulregister\ref7 
\soulregister\pageref7 
\usepackage{bm} 

\usepackage{geometry}
\title{\LARGE \bf
Non-cascaded Control Barrier Functions for the Safe Control of Quadrotors
}

\author{
Weifeng Zeng$^{1}$, Huanhui Cao$^{1}$, Wenjie Lu$^{1}$, and Hao Xiong$^{1,\dag}$
\thanks{$^{1}$Weifeng Zeng, Huanhui Cao, Wenjie Lu, and Hao Xiong are with the School of Mechanical Engineering and Automation, Harbin Institute of Technology Shenzhen, Shenzhen, China.
        }%
\thanks{$\dag$Corresponding author: e-mail: xionghao@hit.edu.cn.}
}

\begin{document}

\maketitle
\thispagestyle{empty}
\pagestyle{empty}

\begin{abstract}
Researchers have developed various cascaded controllers and non-cascaded controllers for the navigation and control of quadrotors in recent years. It is vital to ensure the safety of a quadrotor both in normal state and in abnormal state if a controller tends to make the quadrotor unsafe. To this end, this paper proposes a non-cascaded Control Barrier Function (CBF) for a quadrotor controlled by either cascaded controllers or a non-cascaded controller. Incorporated with a Quadratic Programming (QP), the non-cascaded CBF can simultaneously regulate the magnitude of the total thrust and the torque of the quadrotor determined a controller, so as to ensure the safety of the quadrotor both in normal state and in abnormal state. The non-cascaded CBF establishes a non-conservative forward invariant safe region, in which the controller of a quadrotor is fully or partially effective in the navigation or the pose control of the quadrotor. The non-cascaded CBF is applied to a quadrotor performing trajectory tracking and a quadrotor performing aggressive roll maneuvers in simulations to evaluate the effectiveness of the non-cascaded CBF.

\end{abstract}

\section{Introduction}
Researchers have made great progress with the design, navigation, and control of quadrotors in recent years. 
Several challenging tasks (e.g., surveillance, delivery, and rescue \cite{Wang2017RobustDisturbance,Zhang2012TheReview,Xiong2020Flight-trimArmsb}) have been addressed by quadrotors. 
However, a navigation strategy, a control strategy, or manual manipulations may lead a quadrotor into an unsafe state (e.g., collisions), especially when developing a new navigation strategy or a new control strategy \cite{Xu2018SafeFunctions}.
It’s crucial to ensure the safety of a quadrotor, even if a navigation strategy, a control strategy, or manual manipulations tend to make the quadrotor unsafe.

Control Barrier Functions (CBFs) \cite{Ames2019ControlApplications} have been developed to ensure safety in several applications (e.g., adaptive cruise control \cite{Ames2017ControlSystems}, multi-robot systems \cite{Wang2017SafeFlatness}, carrier landing \cite{Zhou2020ControlLanding}). 
CBFs have been applied to quadrotors by some researchers as well in recent years.
In \cite{Xu2018SafeFunctions}, a position level CBF was designed for assisting human operators to teleoperate quadrotors safely. However, this study is based on the assumption that the Euler angles of a quadrotor are small and cannot be applied to quadrotors performing aggressive maneuvers as a result.
Wang et. al studied the trajectory planning of a team of quadrotors based on differential flatness and CBFs. CBFs were used to modify nominal trajectories for quadrotors to avoid collisions \cite{Wang2017SafeFlatness}.
To address the collision safety constraints suffered by a planar quadrotor, time-varying control Lyapunov functions were used to guarantee stability of the quadrotor and time-varying CBFs were developed to guarantee safety of the quadrotor in \cite{Wu2016Safety-criticalQuadrotor}.
The method was extended to guarantee the safety of a three-dimensional quadrotor in \cite{Wu2016Safety-CriticalSensing}.
Khan et al. proposed cascaded CBFs for a cascaded control architecture of quadrotors to enforce safety, allowing independent safety regulation in the altitude domain and in the lateral domain of a quadrotor \cite{Khan2020BarrierControl}.
However, \cite{Wu2016Safety-criticalQuadrotor,Wu2016Safety-CriticalSensing,Khan2020BarrierControl} focused on quadrotors with cascaded controllers and developed cascaded CBFs for quadrotors. The cascaded CBFs cannot be applied to quadrotors with non-cascaded controllers (e.g., an end-to-end learning-based controller \cite{Zhang2016}). Moreover, the effectiveness of the cascaded CBFs in the case that quadrotors are in abnormal state (e.g., tumbling) has not been verified.






To the best knowledge of the authors, a CBF that can ensure the safety of quadrotors controlled by non-cascaded controllers and can address the safety issue of quadrotors in abnormal state is not available yet.
To address the above-mentioned two issues, this paper proposes a novel non-cascaded CBF for quadrotors. 
The main contributions of this paper are as follows.
\begin{itemize}
    \item This paper develops a non-cascaded CBF for quadrotors controlled by either cascaded controllers or non-cascaded controllers. The non-cascaded CBF can maintain a quadrotor within a safe region in Cartesian space, even if the quadrotor attains an abnormal state.
    \item Based on a physical engine, this paper verifies the effectiveness of the non-cascaded CBF for quadrotors not only in the case of trajectory tracking but also in the case of aggressive roll maneuvers.
\end{itemize}	

The rest of this paper is organized as follows. The preliminaries of this paper are presented in Section II. Section III illustrates a non-cascaded CBF for quadrotors. In Section IV, the non-cascaded CBF is applied to a quadrotor performing trajectory tracking and a quadrotor performing aggressive roll maneuvers based on a physical engine. Section V summarizes this paper.

\section{Preliminary}
This section demonstrates the dynamics of a quadrotor, the problem formulation of the safety of a quadrotor, and exponential control barrier functions.

\subsection{Dynamics of a Quadrotor }
\begin{figure}[htp]
    \vspace{-0.3cm}
    \centering
    \includegraphics[scale=0.09]{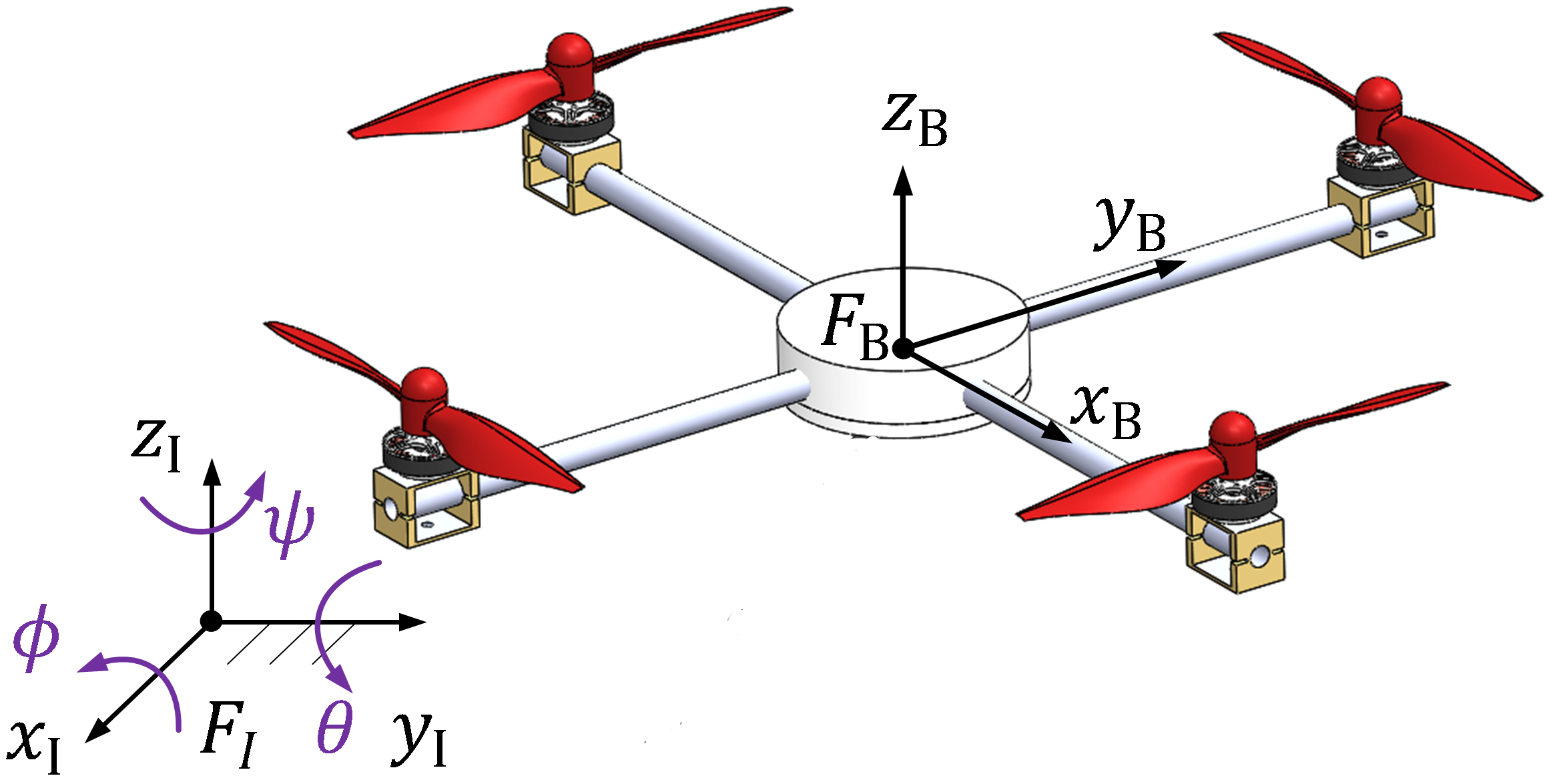}
    \caption{Notations of a quadrotor}
    \label{fig:model}
    \vspace{-0.3cm}
\end{figure}

In this section the dynamics of a quadrotor is demonstrated based on a body frame $F_B$ fixed to the quadrotor and an inertial frame $F_I$, as shown in Fig. \ref{fig:model}. The body frame is defined by axes $x_B$, $y_B$, and $z_B$, and the inertial frame is defined by axes $x_I$, $y_I$, and $z_I$. Euler angles with a Z-Y-X sequence are used to define the roll $\phi$, pitch $\theta$, and yaw $\psi$ angles between the body frame and the inertial frame.

The dynamics of a quadrotor can be expressed as \cite{Zhou2014VectorFlatness}
\begin{equation}
    \begin{cases}
        \dot{\boldsymbol{v}}=-G \bm{z}_I + \frac{1}{m}f_T \bm{z}_B\\
        \dot{\boldsymbol{R}} = \boldsymbol{R} \bm{\Omega}\\
        \boldsymbol{J}\dot{\bm{\omega}}_B = \bm{\tau} - \bm{\omega}_B \times \bm{J} \bm{\omega}_B + \bm{\tau}_g\\
        \dot{\boldsymbol{r}} = \boldsymbol{v}\\
    \end{cases}
    \label{Quadrotor_dynamic}
\end{equation}
where $\boldsymbol{v}=[v_x,~ v_y,~ v_z]^T \in \mathbb{R}^3$ represents the translational velocity of the quadrotor in the inertial frame. 
$G$ is the acceleration of gravity. 
$m$ is the mass of the quadrotor.
$\boldsymbol{z}_I$ is a unit vector in the positive direction of the $z_I$ axis of the inertial frame.
$f_T$ denotes the magnitude of the total thrust of the quadrotor.
$\boldsymbol{z}_B$ is a unit vector in the positive direction of the $z_B$ axis of the body frame.
$\bm{R}$ is the rotation matrix from the body frame $F_B$ to the inertial frame $F_I$ defined in \cite{Zhou2014VectorFlatness}.
$\bm{\omega}_B = [p,~q,~r]^T$ represents the angular velocity of the quadrotor in the body frame.
$\bm{\Omega}$ can be expressed as
\begin{equation}
    \bm{\Omega} =
    \begin{bmatrix}
  0& -r &q \\
 r & 0 & -p\\
 -q & p &0
\end{bmatrix}
\end{equation}
$\bm{J} = diag\{J_{xx},~J_{yy},~J_{zz}\}$ is the inertia matrix of the quadrotor.
$\bm{\tau} = [\tau_x,~\tau_y,~\tau_z]^T$ is the torque generated by the rotors of the quadrotor in the body frame.
$\bm{\tau}_g$ denotes a gyroscopic torque.
$\boldsymbol{r} = [r_x,~r_y,~r_z]^T$ is the position of the quadrotor in the inertial frame. 

The state of a quadrotor can be represented by $\bm{\xi} = [x,~y,~z,~v_x,~ v_y, ~v_z,~ \phi,~ \theta,~ \psi, ~p,~q,~r]^T$.
The control input of the quadrotor is $\bm{u} = [f_T,~ \tau_x,~\tau_y,~\tau_z]^T$. 
The rotor velocities of the quadrotor can be derived based on the control input according to  \cite{Mellinger2011MinimumQuadrotors}

\begin{equation}
    \begin{bmatrix}
    f_T\\ \tau_x\\ \tau_y\\ \tau_z\\
    \end{bmatrix} = 
    \begin{bmatrix}
     k_F&k_F& k_F&k_F & \\
    0 & k_FL&0& -k_FL& \\
     -k_FL& 0& k_FL& 0& \\
     k_M& -k_M& k_M& -k_M& \\
    \end{bmatrix}
    \begin{bmatrix}
    \bar{\omega}_1^2\\ \bar{\omega}_2^2\\ \bar{\omega}_3^2\\ \bar{\omega}_4^2\\
    \end{bmatrix} 
    \label{rotor velocities}
\end{equation}
where $L$ represents the distance between a rotor and the center of the quadrotor. 
$k_F$ and $k_M$ are positive constants determined by air density and the shape of propellers. 
$\bar{\omega}_i$ ($i=1,~2,~3,~4$) denotes the rotational velocity of the $i$th rotor of the quadrotor.

For a quadrotor, one also has \cite{Mellinger2011MinimumQuadrotors}
\begin{equation}
\left\{
    \begin{array}{lll}
        m ~\ddot{\boldsymbol{r}} &=& -mg\bm{z}_I+f_T \bm{z}_B\\
        m\dddot{\boldsymbol{r}} &=& \dot{f_T}\bm{z}_B + f_T\bm{\omega}_{I} \times \bm{z}_B\\
        m\ddddot{\boldsymbol{r}}  &=& \ddot{f_T}\bm{z}_B + 2\dot{f_T} \bm{\omega}_{I} \times \bm{z}_B +\\  & & f_T(\dot{\bm{\omega}_I} \times \bm{z}_B + \bm{\omega}_{I} \times \bm{\omega}_{I}\times \bm{z}_B)
    \end{array}
\right.
    \label{relationship_flat_state}
\end{equation}
where $\bm{\omega}_I = \bm{R}\bm{\omega}_B$ is the angular velocity of the quadrotor in the inertial frame.
In practice, the rotor dynamics are relatively fast for the pose control of a quadrotor, so one can assume that the rotor thrusts as well as the control input of a quadrotor $\bm{u} = [f_T,~ \tau_x,~\tau_y,~\tau_z]^T$ are instantaneously achieved \cite{Mellinger2011MinimumQuadrotors}.
Since quadrotors are usually equipped with digital controllers in practice, given a certain control input, the quadrotor will achieve the control input instantaneously and maintain the control input for a period of time. Thus, one can assume that the derivative of the magnitude of the total thrust $\dot{f_T}$ is zero. 
Then, (\ref{relationship_flat_state}) can be simplified as
\begin{equation}
\left\{
    \begin{array}{lll}
        m ~\ddot{\boldsymbol{r}} &=& -mg\bm{z}_I+f_T \bm{z}_B\\
        m\dddot{\boldsymbol{r}} &=& f_T\bm{\omega}_{I} \times \bm{z_B}\\
        m\ddddot{\boldsymbol{r}} &=& f_T(\bm{\dot{\omega}}_I \times \bm{z}_B + \bm{\omega}_{I} \times \bm{\omega}_{I}\times \bm{z}_B)
    \end{array}
\right.
    \label{relationship_flat_state2}
\end{equation}

\subsection{Problem Formulation}
This paper addresses the safety of a quadrotor in the form of maintaining the quadrotor within a safe region in Cartesian space. 
A safe region is defined as 
\begin{equation}
    \mathcal{S} = \{\boldsymbol{r}\in \mathbb{R}^3 | \boldsymbol{r}_{min} \preceq \boldsymbol{r} \preceq \boldsymbol{r}_{max} \}
    \label{safe area}
\end{equation}
where $\preceq$ represents element-wise inequality. 
$\boldsymbol{r}_{min}$ and $\boldsymbol{r}_{max}$ are the lower bound and the upper bound of $\boldsymbol{r}$, respectively. 
For a quadrotor with the initial position $\boldsymbol{r}(t_0) \in \mathcal{S}$ and a nominal control input in terms of the total thrust and the torque generated by the rotors,
one needs to ensure that the position of the quadrotor satisfies $\boldsymbol{r}(t) \in \mathcal{S}$ where $t \ge t_0$, by regulating the nominal control input.

\subsection{Exponential Control Barrier Function}

Exponential Control Barrier Function (ECBF) is introduced to address the high relative degree safety constraints of a system \cite{Nguyen2016ExponentialConstraints}. The ECBF has been used to design cascaded CBFs for quadrotors in \cite{Khan2020BarrierControl}.
Without loss of generality, one can assume a nonlinear affine system
\begin{equation}
    \dot{x} = f(x) + g(x)u
    \label{nonlinear control affine model}
\end{equation}
where $f$ and $g$ are locally Lipschitz. $x \in D \subset \mathbb{R}^n$ is the state and $u \in U \subset \mathbb{R}^m$ is the control input of the system.
The safety of the system can be guaranteed by enforcing the invariance of a safe set \cite{Ames2019ControlApplications}. In particular, one can consider a set $C$ defined as the superlevel of a continuously differentiable function $h(x):\mathbb{R}^n \rightarrow \mathbb{R}$, yielding
\begin{equation}
\label{math:setC}
\begin{split}
&C=\{x\in \mathbb{R}^n :~ h(x)\ge 0\}\\
&\partial C=\{x\in  \mathbb{R}^n :~ h(x) = 0\}\\
&Int(C)=\{x\in \mathbb{R}^n :~ h(x) > 0\}
\end{split}
\end{equation}
The set $C$ is referred to as a safe set.



\textbf{Definition 1. \cite{Ames2019ControlApplications}} \emph{A set $C$ is forward invariant if for every $x(t_0) \in C$, $x(t) \in C $ for all $t \in [t_0,~ t_{max})$. The system (\ref{nonlinear control affine model}) is safe with respect to the set $C$ if the set $C$ is forward invariant.}

\textbf{Definition 2. \cite{Ames2019ControlApplications}} 
\textit{For system (\ref{nonlinear control affine model}), given a set $C \in D \in \mathbb{R}^n$ defined as the superlevel set of a ${r_b}$-times continuously differentiable function $h(x)$ :~ $D \in \mathbb{R}^n \rightarrow \mathbb{R}$, then $h(x)$ is an ECBF if there exits a row vector $\bm{K}_\alpha \in \mathbb{R}^{r_b}$ such that}
\begin{equation}
\begin{cases}
    &\underset{u \in U}{\sup} [L_f^{r_b}h(x) + L_g L_f^{r_b-1} h(x) u] \ge -\bm{K}_\alpha\bm{\eta}_b(x)\\
    &h(x(t)) \ge \bm{C}_b e^{\bm{A}_bt}\bm{\eta}_b(x_0)\ge 0, \quad h(x(t_0)) \ge 0
\end{cases}
    \label{ECBF}
\end{equation}
\emph{where}
\begin{equation}
    \bm{\eta}_b(x) =
    \begin{bmatrix}
    h(x)\\
    \dot{h}(x)\\
    \ddot{h}(x)\\
    \vdots\\
    h^{r_b-1}(x)\\
    \end{bmatrix} =
    \begin{bmatrix}
    h(x)\\
    L_fh(x)\\
    L_f^2h(x)\\
    \vdots\\
    L_f^{r_b-1}h(x)\\
    \end{bmatrix}
\end{equation}
\begin{equation}
    \bm{C}_b = [1\quad 0 \quad \cdots \quad  0]
\end{equation}
\textit{and the matrix $\bm{A}_b$ is dependent on the choice of $\bm{K}_\alpha$.}

One can define a family of functions $v_i : D \rightarrow \mathbb{R}$ and corresponding superlevel sets $C_i$ ($i = 0,~ \cdots,~ r_b$) as follows \cite{Ames2019ControlApplications}
\begin{equation}
\label{def2}
\left.
    \begin{array}{lll}
        v_0 = h(x),&C_0=\{x:v_0(x)\ge0\},\\
        v_1 = \dot{v}_0(x) + p_1v_0(x),&C_1=\{x:v_1(x)\ge0\},\\
        \vdots&\vdots\\
        v_{r_b} = \dot{v}_{r_b-1}(x) + p_{r_b}v_{r_b-1}(x),&C_{r_b}=\{x:v_{r_b}(x)\ge0\},\\
    \end{array}
\right.
\end{equation}
where $p_i$ is an adjustable parameter and one has $p_i > 0$ ($i =1,~2,~ \cdots, ~r_b$).


\textbf{Theorem 1. \cite{Ames2019ControlApplications}}
\textit{
If $C_{r_b}$ is forward-invariant and $x(t_0)\in \cap^{r_b}_{i=0}C_i$, then $C_0$ is forward-invariant.}

\section{Safe Control of Quadrotors Based on Non-cascaded Control Barrier Functions}
To ensure the safety of a quadrotor controlled by cascaded controllers or a non-cascaded controller,
this section proposes a non-cascaded CBF for the quadrotor.

\subsection{Non-cascaded Control Barrier Functions}
This paper regards a quadrotor as a system with a relative degree of four. Thus, ECBF \cite{Nguyen2016ExponentialConstraints} is used to design non-cascaded CBFs for the quadrotor in this section. 
CBFs are designed for every single $r_x$, $r_y$, or $r_z$ component of the position of the quadrotor, and then
the CBFs are integrated to achieve a CBF that can maintain the quadrotor within a given safe region.

According to the above-mentioned idea, this paper designs two control barrier function candidates for the $r_x$ component of the position of a quadrotor
\begin{equation}
    \begin{cases}
    \overline{h}_x(r_x) = r_{xmax} - r_x\\
    \underline{h}_x(r_x) = r_x - r_{xmin}\\
    \end{cases}
    \label{hx}
\end{equation}
where $r_{xmin}$ and $r_{xmax}$ are the lower bound and upper bound of a safe $r_x$, respectively.
A safe set of the $r_x$ component of the position of a quadrotor is
\begin{equation}
C_x = \{ r_x\in \mathbb{R} | \overline{h}_x(r_x) \ge 0,~ \underline{h}_x(r_x) \ge 0 \}
    \label{mathcal}
\end{equation}

For $\overline{h}_x(r_x)$, a family of functions can be achieved according to \cite{Ames2019ControlApplications}
\begin{equation}
\begin{cases}
\left.
    \begin{array}{lll}
     v_0 =\overline{h}_x \quad& C_0 = \{ r_x | v_0 \ge 0\}, \\
    v_1 =\dot{v}_0 + p_{x_1}v_0 \quad & C_1 = \{ r_x | v_1 \ge 0\},\\
     v_2 =\dot{v}_1 + p_{x_2}v_1 \quad & C_2 = \{ r_x | v_2 \ge 0\},\\
     v_3 =\dot{v}_2 + p_{x_3}v_2 \quad & C_3 = \{ r_x | v_3 \ge 0\},\\
     v_4 =\dot{v}_3 + p_{x_4}v_3 \quad & C_4 = \{ r_x | v_4 \ge 0\},\\
    \end{array}
    \label{confition for x}
\right.
\end{cases}
\end{equation}
where $p_{x_i}$ is an adjustable parameter and one has $p_{x_i}>0$ ($i=1,~2,~3,~4$).
According to (\ref{relationship_flat_state2}) and (\ref{hx}), $v_2$, $v_3$, and $v_4$ depend on the control input of a quadrotor $\bm{u}$. 
If the control input of a quadrotor $\bm{u}$ can enforce the forward-invariance of $C_2,~ C_3$, and $C_4$, the forward-invariance of $C_0$ defined by $\overline{h}_x$ is guaranteed according to Theorem 1. 
One can ensure that $r_x$ is kept a safe distance from $r_{xmax}$ then.

Based on (\ref{hx}) and (\ref{confition for x}), to enforce the forward-invariance of $C_2,~ C_3$, and $C_4$, one has
\begin{equation}
    \boldsymbol{\Gamma}_x \bm{X}\preceq \bm{Q}_x r_{xmax}
    \label{matrixCBFcondition}
\end{equation}
where 
\begin{equation}
\begin{footnotesize}
\begin{cases}
\bm{\Gamma}_x = \begin{bmatrix}
\prod\limits_{i=1}^{2}p_{x_i} & \sum\limits_{i=1}^{2}p_{x_i}  & 1 &  0& 0\\
 \prod\limits_{i=1}^{3}p_{x_i} & \sum\limits_{ 1 \le i< j\le 3}p_{x_i}p_{x_j} & \sum\limits_{i=1}^{3}p_{x_i}  & 1 & 0\\
\prod\limits_{i=1}^{4}p_{x_i}  & \sum\limits_{1 \le i< j < k \le 4}p_{x_i}p_{x_j}p_{x_k} & \sum\limits_{1 \le i< j\le 4}p_{x_i}p_{x_j} & \sum\limits_{i=1}^{4}p_{x_i}  &1
\end{bmatrix} \\
\bm{X} = [r_x,~\dot{r_x},~\ddot{r_x},~\dddot{r_x},~\ddddot{r_x}]^T \\
\bm{Q}_x = [\prod\limits_{i=1}^{2}p_{x_i},~ \prod\limits_{i=1}^{3}p_{x_i},~\prod\limits_{i=1}^{4}p_{x_i} ]^T\\
\end{cases}
\end{footnotesize}
\end{equation}
Similar to (\ref{matrixCBFcondition}), to enforce the forward-invariance of a safe set defined by $\underline{h}_x(r_x)$, one should make
\begin{equation}
    \bm{Q}_x r_{xmin} \preceq \bm{\Gamma}_x\bm{X}     
    \label{minCBFcondition}
\end{equation}
According to (\ref{mathcal}), (\ref{matrixCBFcondition}), and (\ref{minCBFcondition}), one can enforce the forward-invariance of the safe set $C_x$, if
\begin{equation}
    \bm{Q}_x r_{xmin} \preceq \bm{\Gamma}_x\bm{X}\preceq \bm{Q}_x r_{xmax}    \label{X CBFcondition}
\end{equation}
The safety of the $r_x$ component of the position of a quadrotor is ensured then.


One can design CBFs that are similar to (\ref{hx}) for the $r_y$ and $r_z$ components of the position of a quadrotor, according to the approach demonstrated above. 
By integrating the CBFs for the $r_x$, $r_y$, and $r_z$ components of the position of the quadrotor, one can obtain a non-cascaded CBF that can maintain the quadrotor within a safe region. The non-cascaded CBF for the quadrotor can be expressed as
\begin{equation}
    \begin{cases}
    \overline{\bm{h}}= \boldsymbol{r}_{max} - \boldsymbol{r}\\
    \underline{\bm{h}}= \boldsymbol{r} - \boldsymbol{r}_{min} \\
    \end{cases}
\end{equation}
A safe set can be formulated as
\begin{equation}
    C = \{\boldsymbol{r}\in \mathbb{R}^3| \overline{\bm{h}} \succeq 0,~ \underline{\bm{h}} \succeq 0 \}
\end{equation}
To enforce the forward-invariance of the safe set $C$ and ensure the safety of the quadrotor, one should make
\begin{equation}
    \bm{Q}\boldsymbol{r}_{min} \preceq \bm{\Gamma} \bm{\Lambda}\preceq \bm{Q}\boldsymbol{r}_{max}
    \label{safe condition for all}
\end{equation}
where 
\begin{equation}
\begin{footnotesize}
    \begin{cases}
    \bm{\Gamma} = \begin{bmatrix}
        \prod\limits_{i=1}^{2}\bm{P}_i & \sum\limits_{i=1}^{2}{\bm{P}_i}  & \bm{I}_{3\times 3} &   \bm{O}_{3\times 3}&  \bm{O}_{3\times 3}\\
        \prod\limits_{i=1}^{3}{\bm{P}_i} & \sum\limits_{1 \le i < j \le 3}\bm{P}_i\bm{P}_j & \sum\limits_{i=1}^{3}\bm{P}_i  & \bm{I}_{3 \times 3} & \bm{O}_{3 \times 3}\\
        \prod\limits_{i=1}^{4}\bm{P}_i  & \sum\limits_{1 \le i < j < k \le 4}\bm{P}_i\bm{P}_j\bm{P}_k & \sum\limits_{1\le i < j\le 4}\bm{P}_i\bm{P}_j & \sum\limits_{i=1}^{4}\bm{P}_i  &\bm{I}_{3\times 3}
    \end{bmatrix}\\
    \bm{Q} = [\prod\limits_{i=1}^{2}\bm{P}_i,~ \prod\limits_{i=1}^{3}\bm{P}_i,~\prod\limits_{i=1}^{4}\bm{P}_i ]^T\\
    \bm{\Lambda} = [\boldsymbol{r}^T,~\dot{\boldsymbol{r}}^T,~\ddot{\boldsymbol{r}}^T,~ \dddot{\boldsymbol{r}}^T,~\ddddot{\boldsymbol{r}}^T]^T\\
    \bm{P}_i = diag\{p_{x_i}, ~p_{y_i}, ~p_{z_i} \}
    \end{cases}
\end{footnotesize}
    \label{parameters_CBF}
\end{equation}
where $\ddot{\boldsymbol{r}}$, $\dddot{\boldsymbol{r}}$, $\ddddot{\boldsymbol{r}}$ can be adjusted by the control input of the quadrotor according to (\ref{relationship_flat_state2}).

\subsection{Safe Control of a Quadrotor}
According to \cite{Ames2019ControlApplications}, the safe control of a quadrotor can be achieved based on a Quadratic Program (QP) that incorporates CBFs. 
For a nominal control input $\bm{u}_0$ in this paper, a safe control input can be determined by  
\begin{equation}
\left.
    \begin{array}{lll}
    \bm{u}^{*}(t) =& \underset{\bm{u} \in U}{\mathrm{argmin}}\  \|\bm{u}(t) - \bm{u}_0(t)\| \\
    s.t.& \bm{Q}\boldsymbol{r}_{min} \preceq \bm{\Gamma} \bm{\Lambda}(\bm{\xi},~\bm{u}) \preceq \bm{Q}\boldsymbol{r}_{max}\\
    & \bm{u}_{min} \preceq \bm{u} \preceq \bm{u}_{max} \\
    \end{array}
\right.
\end{equation}
where $\bm{u}_{min}$ and $\bm{u}_{max}$ are the lower bound and upper bound of the control input $\bm{u}$, respectively.


In this paper, non-cascaded CBFs are designed for quadrotors with a nominal control input consisting of the magnitude of the total thrust and the torque generated by rotors. However, it should be pointed out that the non-cascaded CBFs can be applied to quadrotors with a nominal control input consisting of rotor thrusts or rotor velocities (e.g., \cite{Zhang2016}) based on (\ref{rotor velocities}).



\section{Simulations}
\begin{figure}[htp]
    \vspace{-0.2cm}
    \centering
    \includegraphics[scale=0.04]{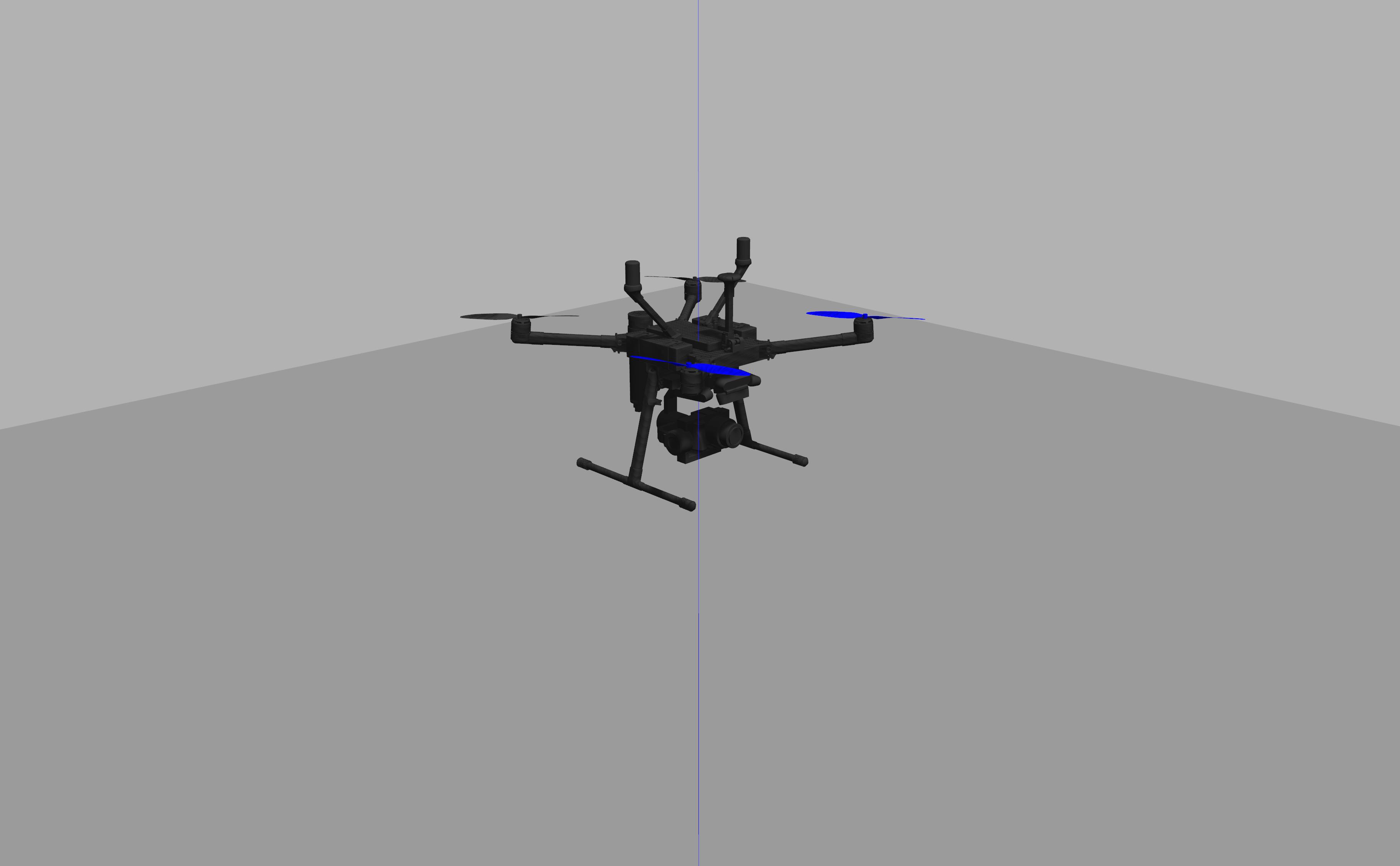}
    \caption{IF750A quadrotor used in simulations}
    \label{fig:if750a}
    \vspace{-0.6cm}
\end{figure}

To evaluate the effectiveness of the developed non-cascaded CBF in ensuring the safety of quadrotors, the non-cascaded CBF is applied to a quadrotor performing trajectory tracking and a quadrotor performing aggressive roll maneuvers in simulations.
The simulations are conducted in a Gazebo simulator \cite{Koenig2004}, a physical engine that has been applied to several studies of quadrotors \cite{Loquercio2020DeepRandomization,Haus2017}. 
An IF750A quadrotor is used in simulations, as shown in Fig. \ref{fig:if750a}. The parameters of the IF750A quadrotor are listed in Table \ref{table1}. $f_{Tmax}$ and $f_{Tmin}$ are the maximum and the minimum feasible magnitude of the total thrust of the quadrotor, respectively. $\tau_{xmax}$, $\tau_{xmin}$, $\tau_{ymax}$, $\tau_{ymin}$, $\tau_{zmax}$, and $\tau_{zmin}$ denote the feasible range of the torque generated by rotors.
\begin{table}[htbp]
	\centering  
	\caption{Parameters of the IF750A quadrotor}  
	\label{table1}  
	\begin{tabular}{|c|c|c|c|}  
		\hline  
		& & &   \\[-6pt]
		Parameter & Value & Parameter & Value   \\
		\hline
		& & &   \\[-6pt]
		$m(\rm{kg})$ & 1.500 & $J_{xx}(\rm{N\cdot m^2})$ & 0.039   \\
		\hline
		& & &   \\[-6pt]  
		 $J_{yy}(\rm{N\cdot m^2})$ &0.051& $J_{zz}(\rm{N\cdot m^2})$ & 0.102   \\ 
		\hline
		& & &  \\[-6pt]
		$f_{Tmax} (\rm{N})$&39.000&$f_{Tmin}(\rm{N})$ & 0.000   \\
		\hline
		& & &   \\[-6pt]
		$\tau_{xmax}(\rm{N\cdot m})$ & 5.130 & $\tau_{xmin}(\rm{N\cdot m})$ & -5.130   \\
		\hline
		& & &   \\[-6pt]
		$\tau_{ymax}(\rm{N\cdot m})$ & 5.130 & $\tau_{ymin}(\rm{N\cdot m})$ & -5.130   \\
		\hline 
		& & &   \\[-6pt]  
	    $\tau_{zmax}(\rm{N\cdot m})$ &  0.024 &$\tau_{zmin}(\rm{N\cdot m})$ & -0.024  \\ 
		 \hline
	\end{tabular}
	\vspace{-0.3cm}
\end{table}

To ensure the safety of the quadrotor in simulations, the parameters of a non-cascaded CBF shown in (\ref{parameters_CBF}) are set to $p_{x_1} = p_{y_1} = 1.000$, $p_{x_2} = p_{y_2} = 4.000$, $p_{x_3} = p_{y_3} = 5.000$, $p_{x_4} = p_{y_4} = 5.000$, $p_{z_1}= 1.000$, $p_{z_2}= 5.000$, $p_{z_3}= 10.000$, $p_{z_4}= 10.000$.

\subsection{Safe Control of a Quadrotor Performing Trajectory Tracking \label{first exper}}

In \cite{Wu2016Safety-criticalQuadrotor,Khan2020BarrierControl}, the effectiveness of cascaded CBFs for quadrotors was validated based on ensuring the safety of a quadrotor performing trajectory tracking.
To evaluate the effectiveness of the developed non-cascaded CBFs, this paper starts by ensuring the safety of a quadrotor performing trajectory tracking as well. 
A safe region is defined as
\begin{equation}
\left.
\begin{array}{lll}
    \mathcal{S} = \{\boldsymbol{r}\in \mathbb{R}^3 |& [-1.000,~-1.000,~2.000]^T \preceq \boldsymbol{r} \\
    & \preceq [1.000,~1.000,~6.000]^T \} ~(\rm{unit: m})
\end{array}
\right.
\end{equation}
A target trajectory that tends to guide the quadrotor to move out of the safe region is defined as
\begin{equation}
\begin{cases}
    &x_{ref} = 0.025~t\cos{(0.200t)} ~(\rm{unit: m})\\
    &y_{ref}= 0.025~t\sin{(0.200t)} ~(\rm{unit: m})\\
    &z_{ref}=3.000+ 0.060t ~(\rm{unit: m})\\
    &\psi_{ref} = 0.000 ~(\rm{unit: rad})
\end{cases}
\end{equation}
Cascaded Proportional-Derivative (PD)-based controllers \cite{Zuo2010TrajectoryMini-helicopter} are used to control the quadrotor to track the target trajectory.
The initial state of the quadrotor is $\bm{\xi} = [0.000,~0.000,~3.000,~0.000,~0.000,~0.000,~0.000,~0.000,\\~0.000,~0.000,~0.000,~0.000]^T$ (units: m, rad, m/s, and rad/s). 

\begin{figure}[htp]
    \vspace{-0.4cm}
    \centering
    \includegraphics[width=8.0cm]{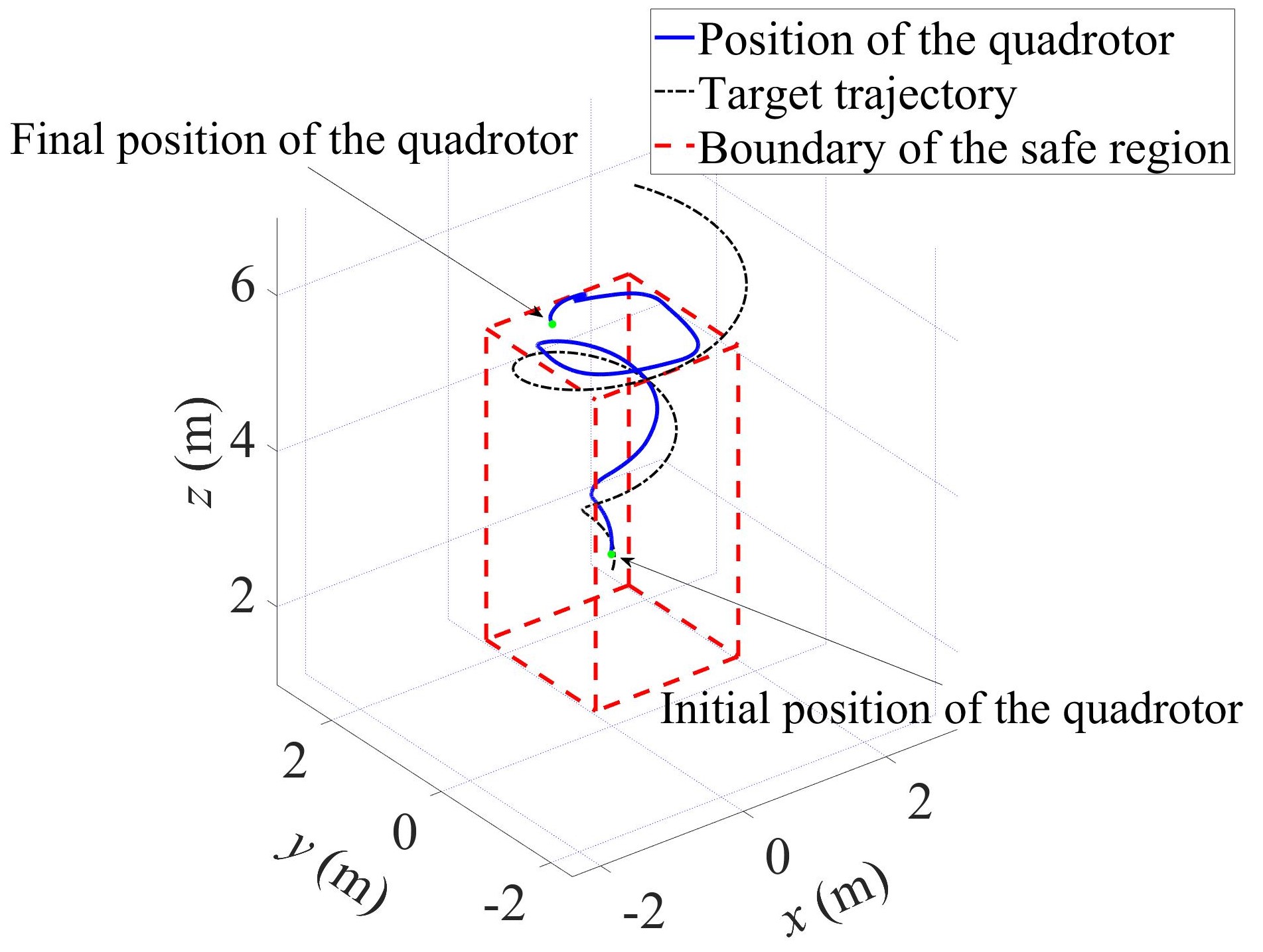}
    \vspace{-0.2cm}
    \caption{Position of the quadrotor performing trajectory tracking}
    \label{fig:3d_position_python}
    \vspace{-0.3cm}
\end{figure}

\begin{figure}[htp]
    \centering
    \includegraphics[width=8.5cm]{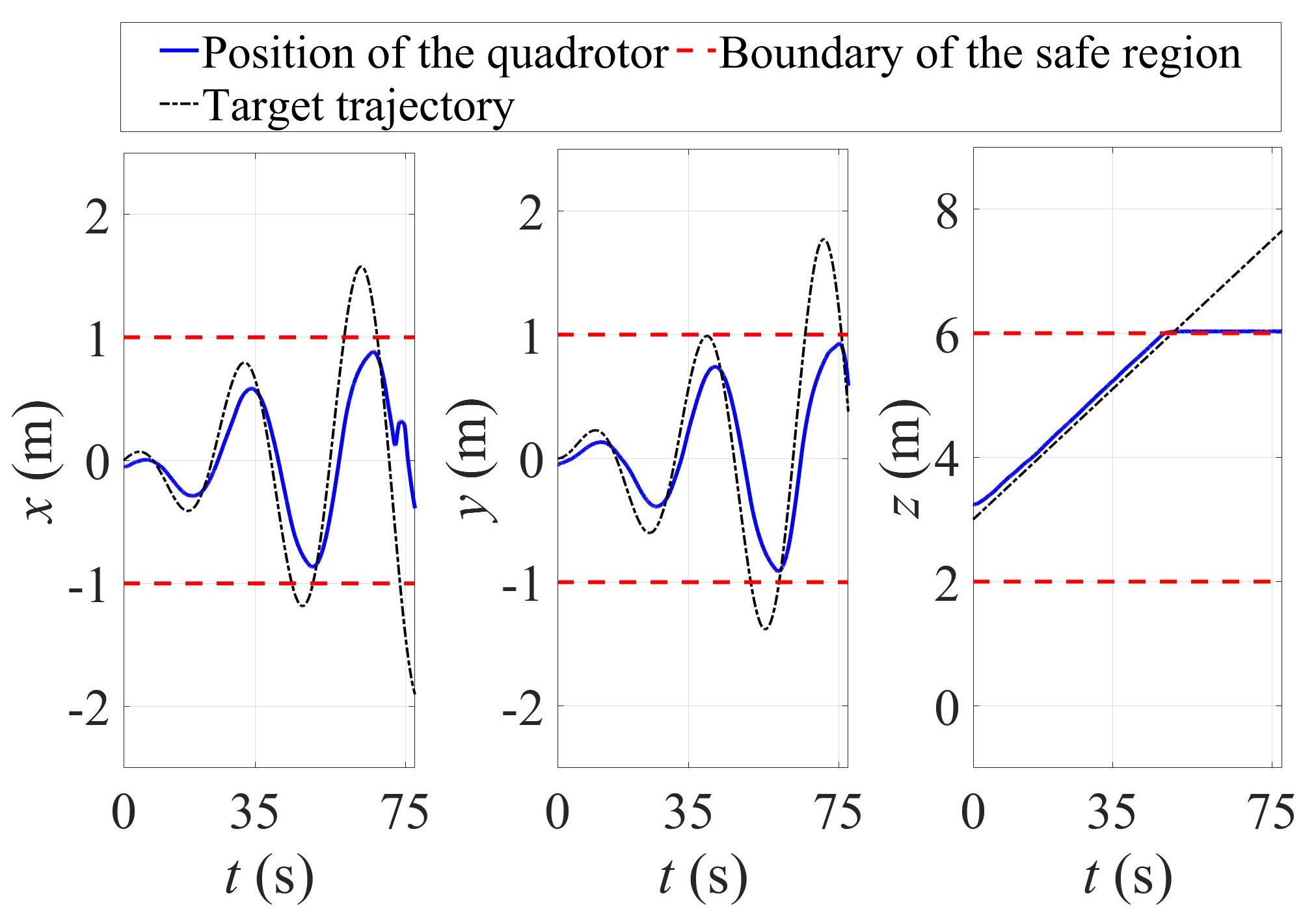}
    \vspace{-0.2cm}
    \caption{Details of the position of the quadrotor performing trajectory tracking}
    \label{fig:position}
    \vspace{-0.3cm}
\end{figure}

The position of the quadrotor performing trajectory tracking is presented in Figs. \ref{fig:3d_position_python} and \ref{fig:position}.
The quadrotor can perform trajectory tracking as long as the quadrotor is within the safe region. 
The trajectory tracking maneuvers of the quadrotor are relaxed when the quadrotor approaches the upper bound of the safe region.
\begin{figure}[htp]
    \centering
    \includegraphics[width= 8.5 cm]{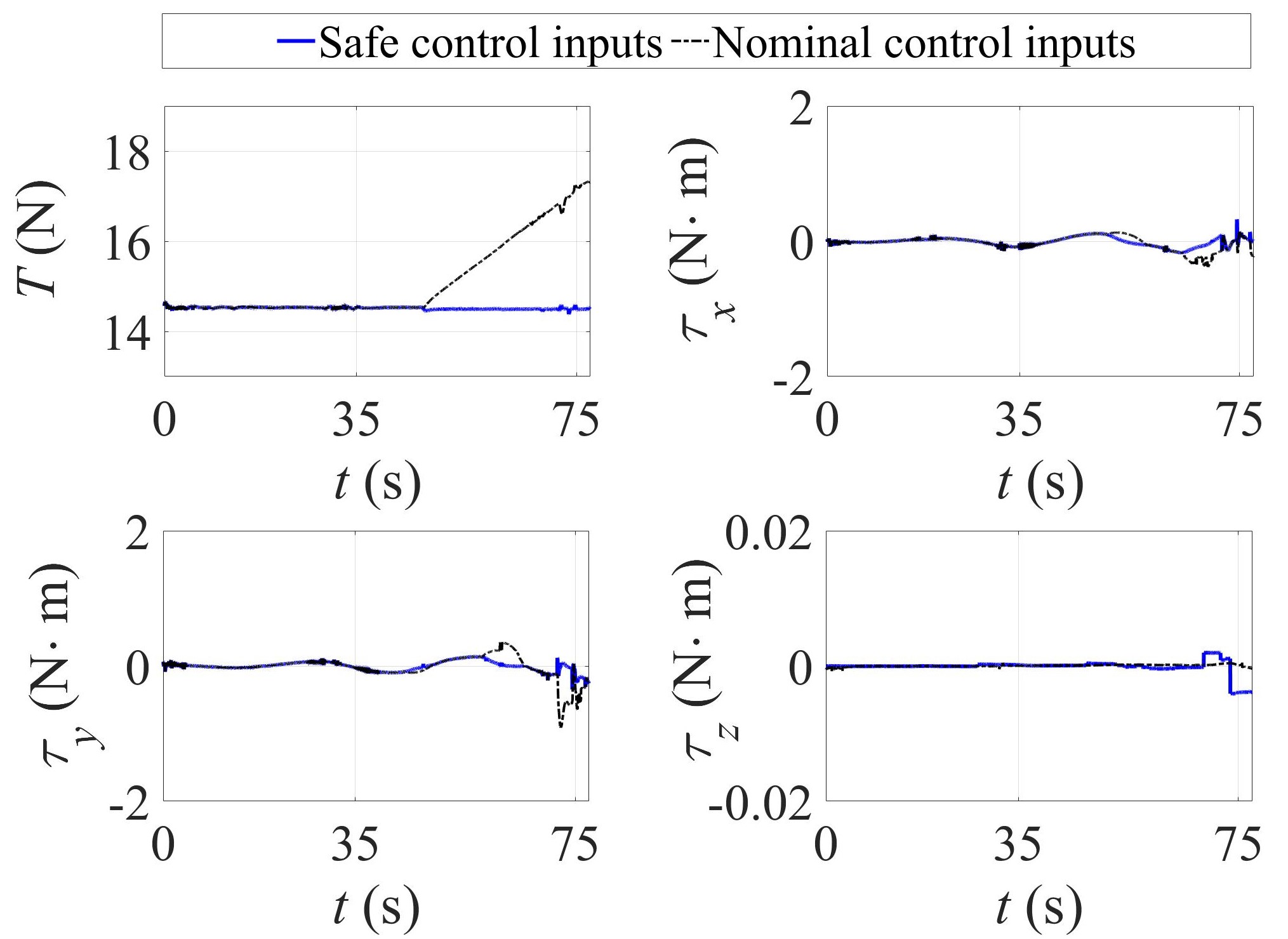}
    \vspace{-0.2cm}
    \caption{Control inputs of the quadrotor performing trajectory tracking}
    \label{fig:input}
    \vspace{-0.7cm}
\end{figure}

To further study the performance of the non-cascaded CBF, the nominal control input and the safe control input of the quadrotor performing trajectory tracking are analyzed. 
Fig.\ref{fig:input} shows that the CBF regulates the nominal control input from about the 45th seconds. This phenomenon is consistent with the position of the quadrotor shown in Fig. \ref{fig:position}. From about the 45th seconds, the nominal control input tends to increase the magnitude of the total thrust of the quadrotor to track the target trajectory, but the CBF regulates the nominal control input to ensure the safety of the quadrotor.
This simulation demonstrates that the non-cascaded CBF is effective in ensuring the safety of a quadrotor performing trajectory tracking.

\subsection{Safe Control of a Quadrotor Performing Aggressive Roll Maneuvers}
A quadrotor may attain an abnormal state in practice, intentionally or unintentionally.
To further evaluate the effectiveness of the non-cascaded CBFs in the case that quadrotors are in abnormal state, this paper applies the non-cascaded CBFs to a quadrotor that tends to perform “Barrel Roll”, an aggressive roll maneuver that can move the quadrotor into abnormal state \cite{Zarchan1979RepresentationFilters}.
A safe region is defined as 
\begin{equation}
\left.
\begin{array}{lll}
    \mathcal{S} = \{\boldsymbol{r}\in \mathbb{R}^3 | &[-4.000,~ -4.000,~ 2.000]^T  \preceq \boldsymbol{r} \\
     & \preceq [4.000,~4.000,~13.000]^T\}~(\rm{unit: m})
\end{array}
\right.
\end{equation}
Assume that a controller gives a nominal control input $\bm{u}_0 = [19.670~\rm{N}, ~0.000~\rm{N\cdot m},~-5.130~\rm{N\cdot m},~ 0.000~\rm{N\cdot m}]^T$ to make the quadrotor to perform the "Barrel Roll" by rolling aggressively.
The initial state of the quadrotor is $\bm{\xi} = [0.000,~0.000,~9.000,~0.000,~0.000,~0.000,~0.000,~0.000, \\~0.000,~0.000,~0.000,~0.000]^T$ (units: m, rad, m/s, and rad/s).

\begin{figure}[htp]
    \vspace{-0.3cm}
    \centering
    \includegraphics[width=8.5 cm]{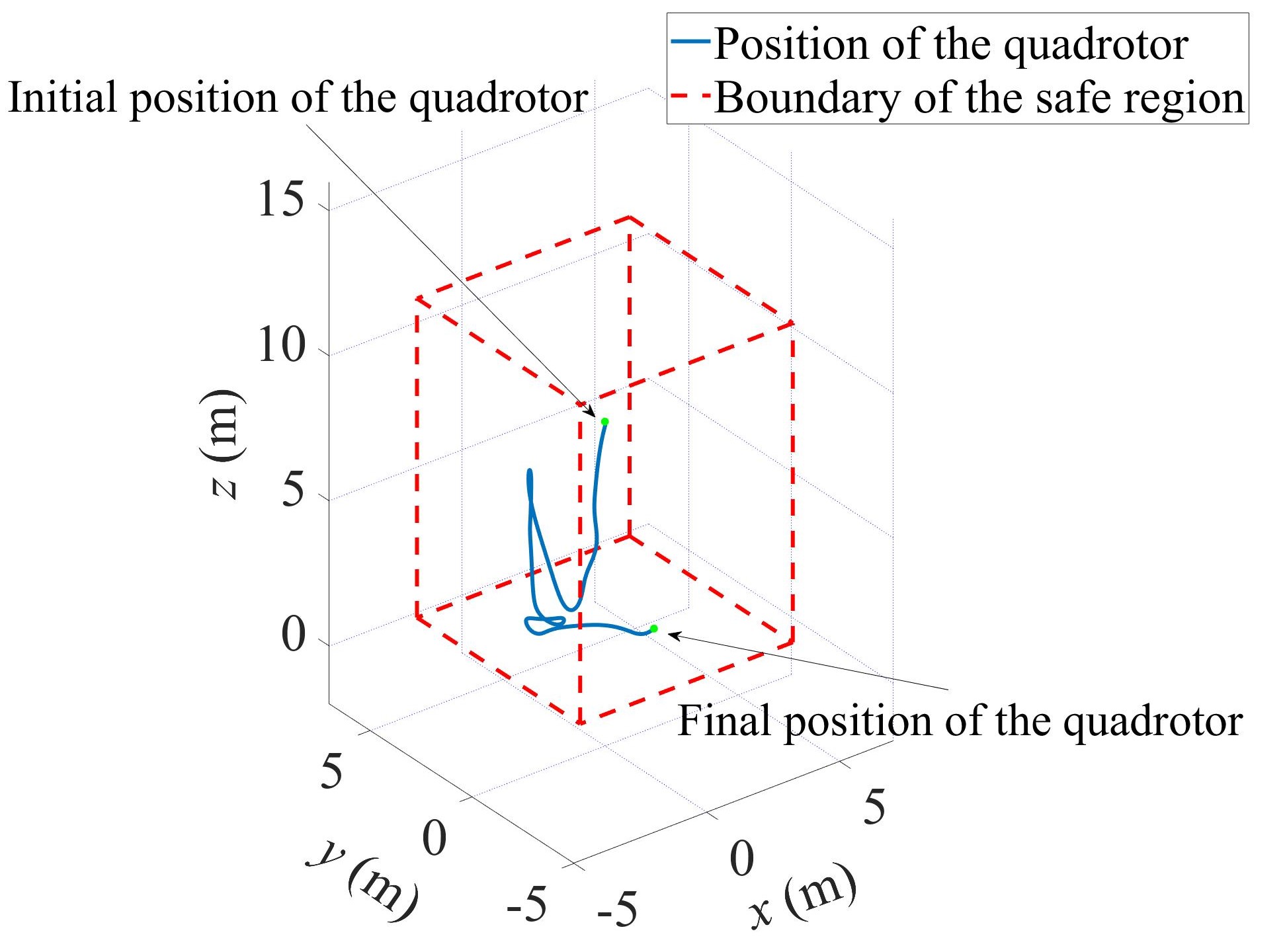}
    \vspace{-0.2cm}
    \caption{Position of the quadrotor performing aggressive roll maneuvers}
    \label{fig:posi_fall_3d}
    \vspace{-0.3cm}
\end{figure}

\begin{figure}
    \centering
    \vspace{-0.5cm}
    \includegraphics[width=8.5cm]{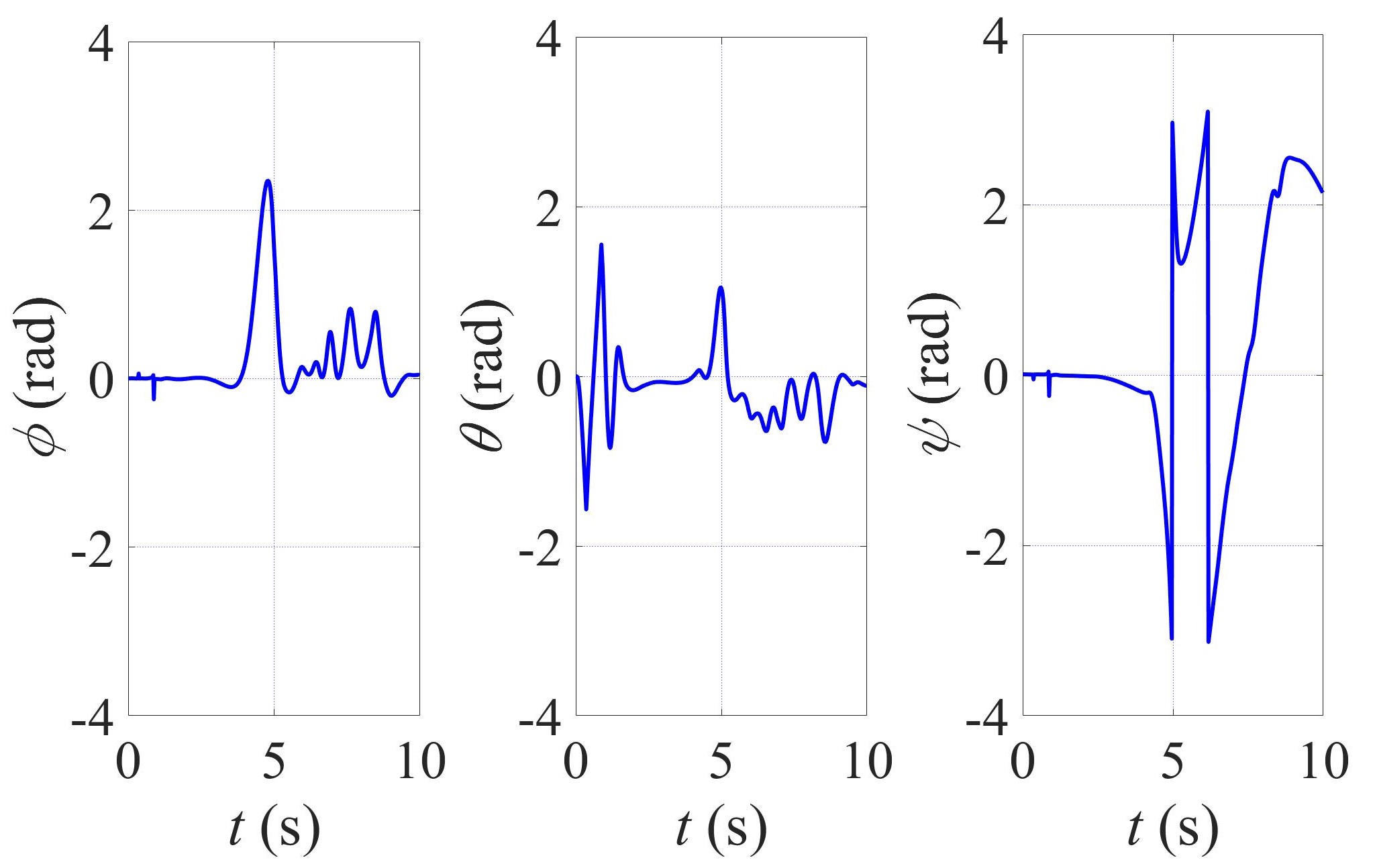}
    \vspace{-0.2cm}
    \caption{Euler angles of the quadrotor performing aggressive roll maneuvers}
    \label{fig:euler_fall}
\end{figure}

\begin{figure}[htp]
    \centering
    \subfigure[]{\includegraphics[height=3.6cm]{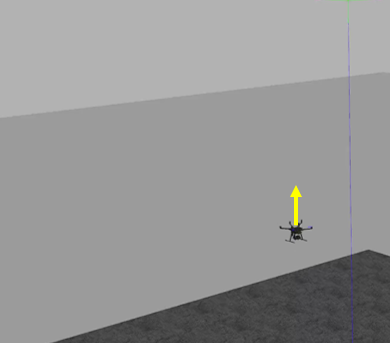}}
    \subfigure[]{\includegraphics[height=3.6cm]{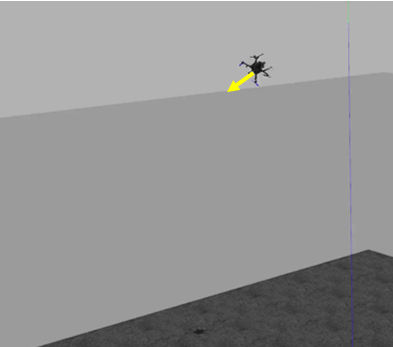}}
    \subfigure[]{\includegraphics[height=3.6cm]{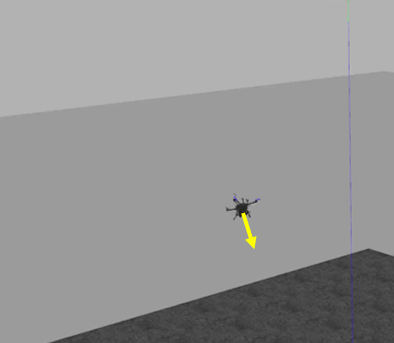}} 
    \subfigure[]{\includegraphics[height=3.6cm]{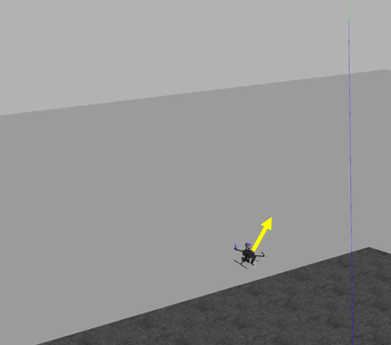}}\\
    \vspace{-0.1cm}
    \caption{Behaviors of the quadrotor performing aggressive roll maneuvers}
    \label{fig:simulation}
    \vspace{-0.3cm}
\end{figure}

The position of the quadrotor performing aggressive roll maneuvers is presented in Fig. \ref{fig:posi_fall_3d}. The non-cascaded CBF can maintain the quadrotor within the safe region indeed, even if the quadrotor tends to roll aggressively.
The orientation and the behaviors of the quadrotor performing aggressive roll maneuvers are shown in Fig. \ref{fig:euler_fall} and Fig. \ref{fig:simulation}, respectively. In Fig. \ref{fig:simulation}, yellow arrows mark the orientation of the $z_B$ axis of the body frame $F_B$. According to \ref{fig:euler_fall} and Fig. \ref{fig:simulation}, one can see that the quadrotor has rolled several times and attains an abnormal state several times. The non-cascaded CBF can maintain the quadrotor within the safe region, even if the quadrotor is in abnormal state.

\begin{figure}[htp]
    \centering
    \includegraphics[width=8.6cm]{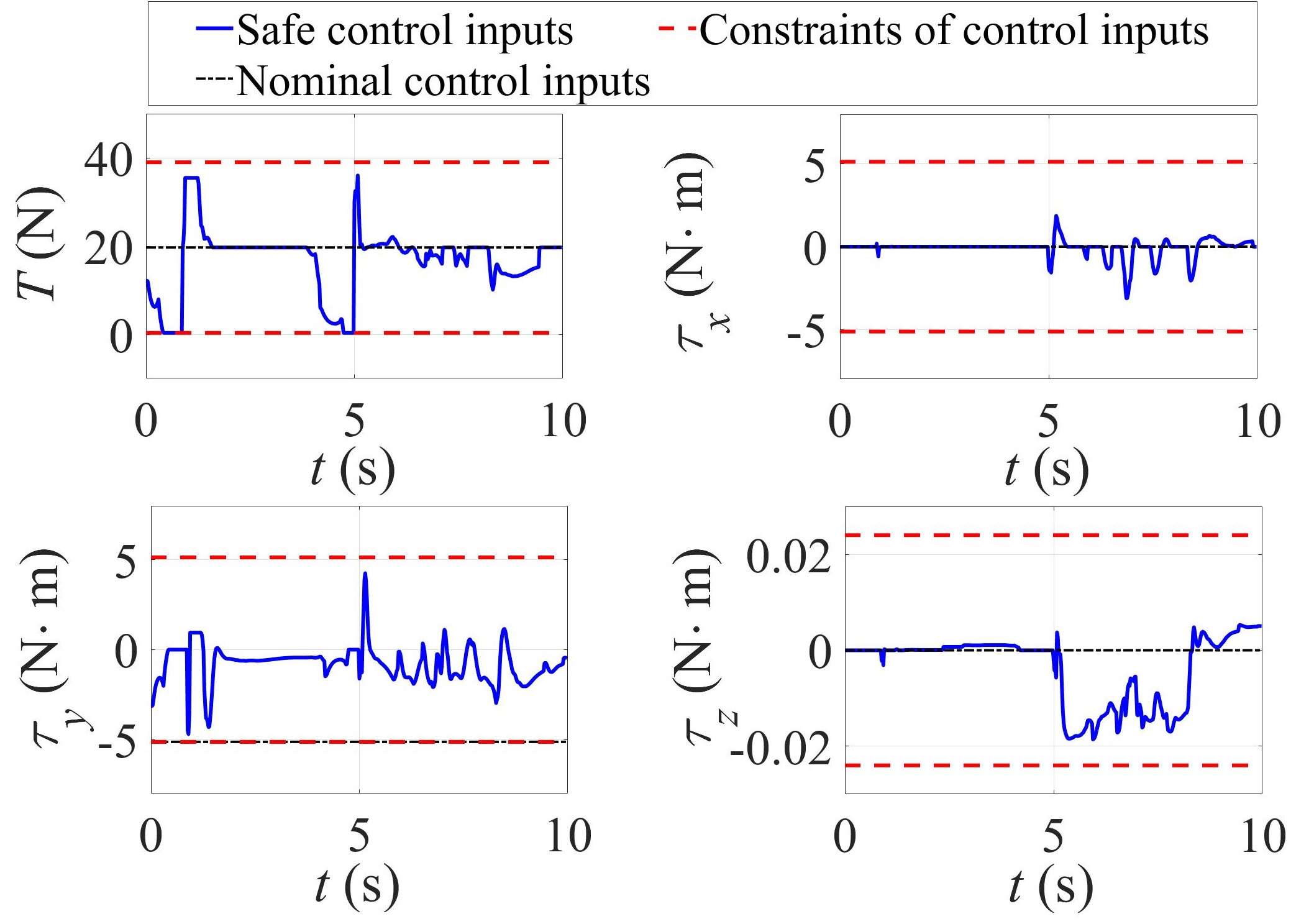}
    \vspace{-0.2cm}
    \caption{Control inputs of the quadrotor performing aggressive roll maneuvers}
    \label{fig:input_fall}
    \vspace{-0.7cm}
\end{figure}

To further study the performance of the non-cascaded CBF, this paper investigates the nominal control input and the safe control input of the quadrotor performing aggressive roll maneuvers, as shown in Fig.\ref{fig:input_fall}.
The non-cascaded CBF regulates the nominal control input of the quadrotor during the whole simulation to maintain the quadrotor within the safe region. Although the non-cascaded CBF enforces the quadrotor to relax aggressive roll maneuvers to uphold safety, the quadrotor has finished the "Barrel Roll" according to Fig \ref{fig:simulation}. This suggests that the non-cascaded CBF regulates rather than takes over the control input. Thus, the non-cascaded CBF enables the quadrotor to perform maneuvers and tasks, to some extent, while enforcing the safety of the quadrotor.




\section{Conclusions}

This paper developed a non-cascaded CBF for quadrotors that may attain abnormal states controlled by cascaded controllers or non-cascaded controllers.
Based on the dynamics of a quadrotor, ECBF was used to design a non-cascaded CBF for quadrotors with a nominal control consisting of the magnitude of the total thrust and the torque generated by rotors.
The non-cascaded CBF has been applied to a quadrotor performing trajectory tracking and a quadrotor performing aggressive roll maneuvers to ensure the safety of the quadrotors in simulations.
The results of the simulations verify the effectiveness of the non-cascaded CBF, even if a quadrotor is in abnormal state.





\bibliographystyle{IEEEtran}
\bibliography{references}

\end{document}